\documentclass[letterpaper]{article} 
\usepackage{aaai23-r2hcai}  
\usepackage{times}  
\usepackage{amsmath}
\usepackage{helvet}  
\usepackage{courier}  
\usepackage[hyphens]{url}  
\usepackage{graphicx} 
\usepackage{natbib}  
\usepackage{caption} 
\usepackage{algorithm}
\usepackage{algorithmic}

\usepackage{newfloat}
\usepackage{listings}
\DeclareCaptionStyle{ruled}{labelfont=normalfont,labelsep=colon,strut=off} 
\floatstyle{ruled}
\newfloat{listing}{tb}{lst}{}
\floatname{listing}{Listing}
\usepackage{fancyhdr}
\fancypagestyle{firstpagehf}
{
   \fancyhf{}
   \fancyhead[HC]{Published in PNAS Nexus}
   \fancyfoot[C]{\thepage}
}

\title{Complex Systems of Secrecy: 
The Offshore Networks of Oligarchs}
\author{
    Ho-Chun Herbert Chang$^1$ \quad
    Brooke Harrington$^2$ \quad
    Feng Fu$^3$ \quad
    Daniel Rockmore$^{3,4}$
}
\affiliations {
    \textsuperscript{\rm 1} Department of Quantitative Social Science, Dartmouth College Hanover NH, 03755  \quad
    \textsuperscript{\rm 2} Department of Sociology, Dartmouth College Hanover NH, 03755 \quad
    \textsuperscript{\rm 3} Department of Mathematics, Dartmouth College Hanover NH, 03755 \quad
    \textsuperscript{\rm 4} Santa Fe Institute, Santa Fe NM, 87501 \\
    herbert@dartmouth.edu
}

\usepackage{bibentry}

\begin{document}
\thispagestyle{firstpagehf}
\maketitle

\begin{abstract}
Following the invasion of Ukraine, the US, UK, and EU governments--among others--sanctioned oligarchs close to Putin. This approach has come under scrutiny, as evidence has emerged of the oligarchs' successful evasion of these punishments. To address this problem, we analyze the role of an overlooked but highly influential group: the secretive professional intermediaries who create and administer the oligarchs' offshore financial empires. Drawing on the Offshore Leaks Database provided by the International Consortium of Investigative Journalists (ICIJ), we examine the ties linking offshore expert advisors (lawyers, accountants, and other wealth management professionals) to ultra-high-net-worth individuals from four countries: Russia, China, the United States, and Hong Kong. 
We find that resulting nation-level “oligarch networks” share a scale-free structure characterized by a heterogeneity of  heavy-tailed degree distributions of wealth managers; however, network topologies diverge across clients from democratic versus autocratic regimes.  While generally robust, scale-free networks are fragile when targeted by attacks on highly-connected nodes. Our “knock-out” experiments pinpoint this vulnerability to the small group of wealth managers themselves, suggesting that sanctioning these professional intermediaries may be more effective and efficient in disrupting dark finance flows than sanctions on their wealthy clients. This vulnerability is especially pronounced amongst Russian oligarchs, who concentrate their offshore business in a handful of boutique wealth management firms. 
The distinctive patterns we identify suggest a new approach to sanctions, focused on expert intermediaries to disrupt the finances and alliances of their wealthy clients. More generally, our research contributes to the larger body of work on complexity science and the structures of secrecy.
\end{abstract}

\section{Introduction}

Russia’s invasion of Ukraine in February 2022 was swiftly met with sanctions imposed by nations around the world, including the US, the UK, Canada, Australia, Switzerland, Singapore, the European Union, and its member states. Among the many sanctions regimes put in place, those targeted at individual oligarchs close to Putin garnered particular attention. Informed by economic research showing that Russian elites hold a staggering 60 percent of their wealth offshore—compared to the 10 percent global average for the ultra-wealthy~\cite{Als2018taxhavens}—these sanctions yielded dramatic media coverage of yacht and luxury villa seizures. 

But regulatory reviews have exposed major gaps and “shortcomings”~\cite{Wolcott2022Reuters}
in these sanctions regimes. Shortly after they were imposed, evidence emerged showing some prominent Russians evading sanctions through the assistance of elite intermediaries: trusted experts who secretly transferred the oligarchs’ financial assets~\cite{Gregory2022BBC} and spirited away treasures like priceless art collections beyond the reach of the law. The same happened following efforts to sanction Russian oligarchs after the 2014 invasion of Crimea~\cite{Bertrand2022Politico}. Oligarchs' ability to escape  sanctions persists because governments have been “slow to address the enabler problem,” in part due to the exorbitant money and time required to investigate the secretive mechanisms used by those intermediaries. 

Any efforts to sanction or curb elites’ accumulation and abuses of wealth face the challenge of secrecy. It is an old problem: the 19th-century sociologist Georg Simmel observed that secrecy links all groups seeking wealth in order to escape from the rule of law, whether they be members of the nobility, bandit gangs, or other “predatory" associations~\cite{simmel1906sociology}. In the contemporary context of offshore finance, secrecy is the main product sold to elites~\cite{harrington2012trust, harrington2016capital,hoang2022spiderweb,winters2011oligarchy}. The result is what anthropologist Bill Maurer calls “non-locatable structures of domination” –a strategic obscurity surrounding who owns which assets, making it all but impossible to tax, sanction, or otherwise hold elites accountable~\cite{maurer1995complex}. This creates considerable uncertainty about whom to target with policy measures, and how. 

In this paper, we introduce a network approach to the study of oligarch sanctions, with implications for both scholarship and policy-making. Our analysis offers an efficient alternative strategy for achieving policy objectives targeting the offshore wealth of oligarchs, particularly those linked to authoritarian regimes. Using the Offshore Leaks Database provided by the International Consortium of Investigative Journalists (ICIJ) we create and analyze the global offshore networks linking high-net-worth individuals to the professional intermediaries who manage their offshore wealth. We focus on elite clients from four countries: Russia, China, Hong Kong, and the United States. While our primary interest is in the recently-sanctioned Russian elites, this larger group of countries provides a useful general context for our findings. Recent economic analyses identify these four nations as the most economically unequal in the world, with the greatest concentration of wealth in the .001\% of the population: the realm of oligarchs~\cite{piketty2021income}.

Our analysis of these “oligarch networks” provides the first quantitative test of two key implications from prior ethnographic research~\cite{grigoropoulou2022data} on the offshore fortunes of high-net-worth individuals. This group includes oligarchs, defined as ultra-wealthy people who use their fortunes to exert a disproportionate influence on politics—domestic and international. By building on in-depth ethnography with network analysis using “big data” from offshore leaks, our investigation answers recent calls to leverage synergies between qualitative and quantitative research~\cite{grigoropoulou2022data}.

The first implication we examine from earlier qualitative studies is that intermediary wealth managers—rather than the elites they serve—are the linchpins binding the global system of offshore finance, without whom the system could not function~\cite{harrington2016capital}. The second implication we study is that, despite their many commonalities, there remain key distinctions among oligarchs from different countries, particularly in the patterns of their relationships with wealth managers~\cite{harrington2018between,harrington2016capital}.

Our first discovery is that oligarch networks have a “scale-free structure,” meaning that they are degree-heterogeneous networks with heavy-tailed degree distributions. Oligarch networks thus possess the characteristic structural vulnerability of scale-free networks: they can be disrupted most efficiently through targeted attacks manifested as the deletion of a few highly connected nodes~\cite{albert2000error}. This is the so-called, “robust yet fragile” property of scale-free networks, a term coined by Albert, Barabasi, and Jeong ~\cite{albert2000error} and characteristic of many networks, including the Internet~\cite{doyle2005robust}. We identify financial intermediaries—wealth management professionals such as the lawyers, accountants, bankers and others who specialize in serving the ultra-rich—as a class of highly connected nodes who constitute an overlooked source of fragility in the global and nation-level offshore financial networks. This suggests a surprising result: the most effective and efficient way to punish oligarchs may be to sanction their offshore intermediaries, the wealth managers. 

These networks can be robust yet “super-fragile”, which is to say that (as shown in our series of “knockout" experiments) deleting a very small number of intermediaries can effect significant structural damage to some nation-level oligarch networks. Furthermore, fine-scale structural differences appear at the national level, reflected in patterns of relational homophily. These fine-grained differences across scale-free networks—which we link to divergent political systems and historical path dependencies—constitute our second discovery. Our finding extends Simmel’s work on the limits of trust among elites linked by secrets, and aligns with qualitative research indicating high levels of distrust among high-net-worth individuals~\cite{sherman2019uneasy}. This distrust can extend to the point of paranoia in some cases, particularly when it comes to sharing information about the sources and locations of their wealth—even when dealing with the professional wealth managers who most need to know~\cite{harrington2016capital,harrington2018between,marcus_hall_1992}. The significance of secrecy and distrust appears to increase in importance under two conditions characteristic of autocracies: corruption and lack of confidence in institutions such as the rule of law~\cite{gambetta2000mafia}. Hence, qualitative research suggests that oligarchs from non-democratic regimes like Russia tightly restrict their circle of advisors, doling out information only on a “need to know” basis and exclusively to a handful of trusted experts—usually chosen by word of mouth from family and friends, rather than by shopping around for the most knowledgeable or  -effective provider~\cite{harrington2016capital}. 

The relational patterns we discover in these financial networks are effectively networks  of secrecy and are shaped in part by historical path dependencies—notably the legacy of British colonialism, which created the legal basis for most of the offshore world~\cite{ogle2017archipelago}. More specifically, since British Common Law is the “operating system” of offshore finance, access to the system is easier for individuals whose native countries share those legal institutions and the English language~\cite{harrington2016capital,palan2002tax}. These historical path dependencies make it easier for wealthy individuals from the US and Hong Kong—themselves former British colonies—to obtain expert offshore advisory services close to home. In contrast, the system is less accessible to elites from places that were never colonized and don’t share the language or legal institutions that are dominant offshore~\cite{chavagneux2013tax,ogle2017archipelago}. This may explain the distinctive patterns of Chinese and Russian engagement with the offshore financial system. For Chinese high-net-worth individuals, Hong Kong is their primary access point to expert wealth managers providing offshore services; as a former British colony and one of the world’s most established offshore financial centers, Hong Kong is replete with professionals who speak Chinese, understand Chinese culture and politics, and therefore can serve Chinese clients better than wealth managers from outside the region. Since Russian oligarchs have no equivalent to Hong Kong—that is, no offshore financial center on the borders of their country—these individuals instead access the system through experts based in a diverse range of countries, many of which are former British colonies.  We quantify these differences in relationship diversity through a novel application of the “Herfindahl-Hirschman Index” (HHI), a well-known diversification metric of market concentration~\cite{rhoades1993herfindahl}.

More generally, our work fits into the broader body of network science research that has been instrumental in unveiling secretive, deceptive, and even criminal behavior. For instance, a network approach has shed light on the ways criminal alliances can be leveraged for more effective law enforcement~\cite{klerks2004network,duijn2014social,mcillwain1999organized}.
Other examples include the study of the structure of state-controlled attacks on social media during elections~\cite{chang2021social,chang2021digital,ferrara2020characterizing}, on the communication patterns within terrorist organizations~\cite{matusitz2008similarities}, and on price-fixing conspiracies~\cite{baker1993social,enders2007rational}. Research on networks of illegal activity has identified brokers--the operational coordinators of grand larceny and other covert offenses--as the linchpins of criminal enterprises;  their removal devastates the network overall~\cite{bright2017criminal,morselli2008brokerage,morselli2010assessing}. These brokers are characterized by high betweenness centrality, which indicates how crucial they are to facilitating shortest paths. However, only a handful of studies~\cite{dominguez2020panama,poon2019social} have applied this analytic technique to the study of offshore wealth management. None as yet have conducted systematic assessments of the expert networks' robustness in the face of disruptions such as sanctions. This is our distinctive contribution, realized as the deletion of certain expert nodes in the offshore financial networks of oligarchs.

Our investigation and the research cited above can be viewed as a part of a growing body of work on what we call the “complex systems of secrecy.” The scale-free structure we uncover in oligarchs’ offshore networks, particularly their overlooked vulnerability to attack via removal of central wealth managers with dense ties, exemplify a phenomenon identified in new research on “complex secrets”~\cite{rilinger2019corporate}. This novel theory of secrecy builds on Simmel’s work to posit that the accumulation of wealth and power by elites can be most effectively achieved when key actors and data necessary to maintaining secrets are distributed across multiple—often unlikely—locations in a network. We extend the implications of this theory by showing that wealth managers represent these overlooked repositories of secrets within the complex networks of oligarchs’ offshore fortunes. This network framing, characterization of roles in terms of network position, and distributed structure suggest that our findings and methods can establish a basis for more effective policy interventions, and for more fruitful research on secrecy as a complex system. 

\section{Results}

\begin{figure*}[!hbt]
    \centering
    \includegraphics[width=0.7\linewidth]{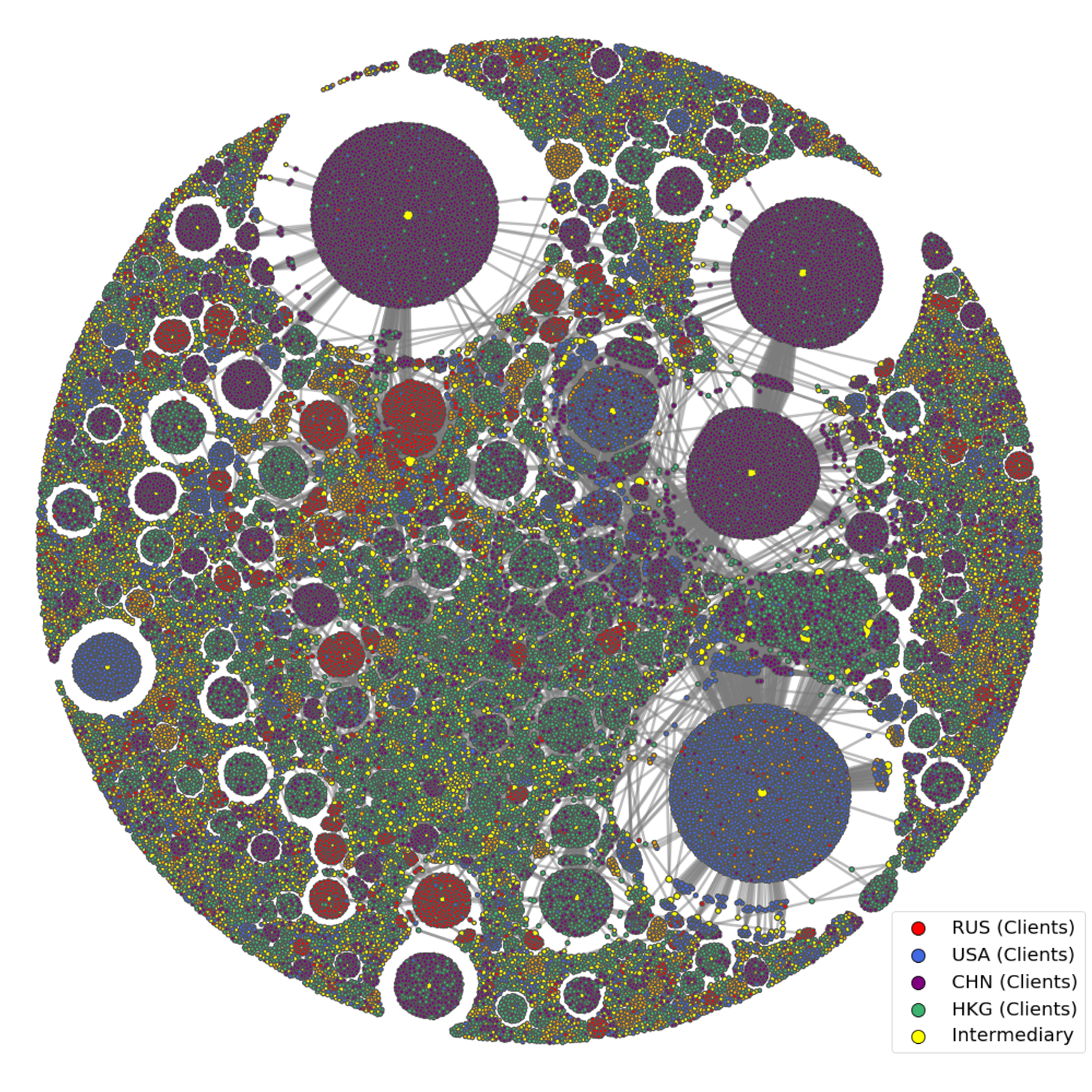}
    \caption{Offshore bipartite financial networks constructed using the ICIJ offshore leaks database. For clarity, shown here is the partial network of 79,458 intermediaries and their clients from Russia (RUS/red), China (CHN/purple), Hong Kong (HKG/green), and the United States (USA/blue). Nodes are either beneficiaries or intermediaries, visualized through physics-based verlet integration. This is a subset of a greater network of 1,970,448 nodes and 3,273,524 edges.}
    \label{fig:network_viz}
\end{figure*}

Our analysis focuses on the network of relationships connecting clients (also known as beneficiaries), wealth managers (also known as intermediaries and offshore entities (such as corporate service providers).\footnote{We use the terms "beneficiary," "intermediary," and "entities" in keeping with the norms of the financial services industry~\cite{boyd1986financial,hoang2022spiderweb}, as reflected in the terminology found in the ICIJ dataset.  We emphasize that none of the terms are technical social network measures; for example, "intermediary" does not connote centrality in an arbitrary network. That said, the oligarch networks are bipartite (connecting intermediaries, clients, and offshore entities) and in this case, intermediaries do in fact lie between clients who are not connected to each other.} In total, the data comprise 1,970,448 such entities (the nodes in the network), along with 3,273,524 relationships: edges connecting intermediaries, clients, and entities. From that superset, we extract a subset of interest, consisting of ultra-high-net worth individuals from countries known to produce oligarchs—Russia, China, Hong Kong and the United States—for a total of 79,458 clients and 143,795 edges. Figure~\ref{fig:network_viz} visualizes the network of intermediaries and 90,155 of their clients from Russia (RUS), China (CHN), Hong Kong (HKG), and the United States (USA) (see data and methods). Centered around these clusters are intermediaries, marked in yellow.

\begin{table}[!htb]
\begin{tabular}{|l|l|l|l|l|l|}
\hline
\textbf{Country} & \textbf{Clients} & \textbf{\begin{tabular}[c]{@{}l@{}}Inter-\\ med\end{tabular}} & \textbf{Edges} & \textbf{LGC Size} & \textbf{\% GC} \\ \hline
CHN              & 32,045           & 1,601                                                         & 48,239         & 22,235           & 66.1\%          \\ \hline
RUS              & 6,311            & 510                                                           & 8,512          & 4,267            & 62.6\%           \\ \hline
USA              & 15,450           & 1,632                                                         & 32,253         & 8,647            & 50.6\%           \\ \hline
HKG              & 25,661           & 3,665                                                         & 54,791         & 15,785           & 53.8\%           \\ \hline
All 4            & 79,458           & 5,711                                                         & 143,795        & 69,916           & 60.1\%           \\ \hline
\end{tabular} \caption{Oligarch network statistics. The rightmost column, “\% GC” counts the proportion of the largest connected component (LGC) in each network. Note that they are each dominated by a large “giant component” (GC). } \label{tab:network_stats}
\end{table}

We further partition this aggregate network into four nation-level oligarch networks, one each for Russia, China, Hong Kong, and the United States. Each nation-level network is constructed by first extracting all beneficiaries from a specific country. We then identify entities they are associated with, and the intermediaries who created the entities, who are the intermediaries included in the network. The characteristics of the nation-level networks are given in Table~\ref{tab:network_stats}.\footnote{We use the terms “intermediaries” and “wealth managers” interchangeably, and the terms “clients” and “beneficiaries” interchangeably. Both the terms “intermediaries” and “wealth managers” signify experts whose formal training is in the law, accountancy, banking, tax advisory services and related fields. By the same token, the word “beneficiaries” defines “clients” in relation to the legal structures that wealth managers create—as in the phrase “beneficiaries of offshore trusts.” In our case-study, we further define “oligarchs” as a subset of ultra-high-net-worth individuals who devote some of their time and wealth to influence affairs of state, whether domestic or international.} 

\subsection{Country-level Comparisons}
\begin{figure}[!hbt]
    \centering
    \includegraphics[width=0.8\linewidth]{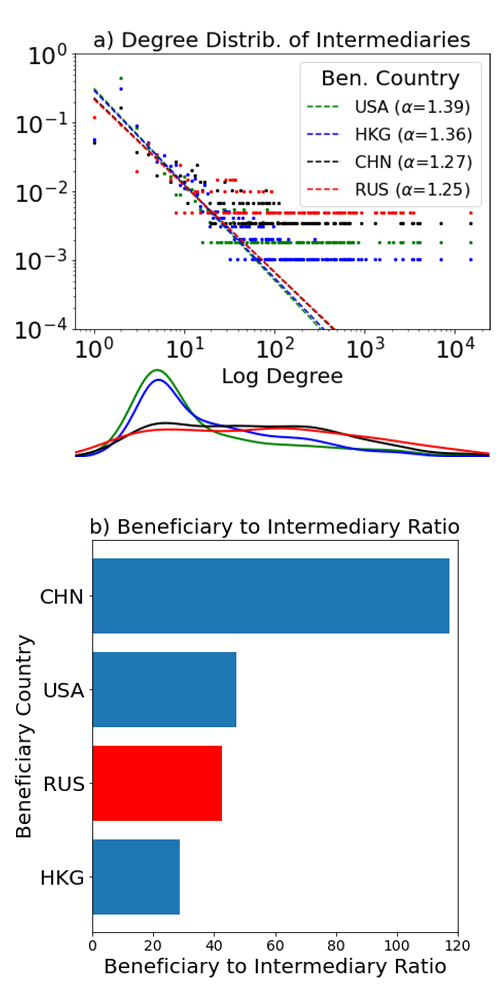}
    \caption{A Matthew Effect of trust in accumulated advantage for offshore wealth managers. A small fraction of these experts serve disproportionately large numbers of clients. Shown are degree properties by clients of certain countries. Figure~\ref{fig:powerlaw}(a) shows the degree distribution of these intermediaries. Figure~\ref{fig:powerlaw}(b) shows the ratio of beneficiaries to wealth managers/intermediaries.}
    \label{fig:powerlaw}
\end{figure}

Table~\ref{tab:network_stats} summarizes the characteristics of the nation-level oligarch networks. All networks are bipartite, connecting clients  with intermediaries; while highly disconnected, they are dominated by a sizable giant (connected) component. Both the aggregate network and the nation-level networks have scale-free degree distributions (see Figure S2 in the SI). Clients are disproportionately represented in the spike at the small degree, reflecting a strategy of trusting just a handful of intermediaries, with an average degree of 1.036 across the clients of all countries. Figure~\ref{fig:powerlaw}(a) isolates the degree distributions of intermediaries for clients based in the US, Hong Kong, China, and Russia respectively. Each are well-fit by a power law distribution (see test statistics in Supplementary Information). This is consistent with—although not required for—a growth process in which one person’s choice of wealth manager (which is a signal of trust) inspires others to trust in the same intermediary. This preferential attachment process creates a “Matthew Effect” for the intermediaries.  

While the slopes of their fits are close, there is a marked distinction between the distribution patterns for Russia and China versus those for the United States and Hong Kong (full robustness checks of the powerlaw fit can be found in Table S3 of the SI). Indeed, when we consider the kernel density estimates below, we find the curves for China (black) and Russia (red) overlapping and much flatter than those of the United States (green) and Hong Kong (blue). Together, this suggests divergent macroscale behavior between autocratic and democratic nations. This may be because oligarchs perceive wealth managers with fewer clients as being at lower risk of breaching secrecy, simply because there aren’t as many people interacting with the professionals—hence, there are fewer points of potential data leakage. Previous research on Mafia networks suggests that this is a particular concern among clients linked to corruption or criminality~\cite{gambetta2000mafia}.

However, Figure~\ref{fig:powerlaw}(b) shows how the degree distribution may not necessarily correspond to the ratio between beneficiaries and intermediaries. China has by far the largest number of clients per intermediary, followed by the United States, then Russia and Hong Kong. We attribute this variation to supply and demand for wealth management services. For example, while China now has more billionaires than any other country, and is producing new billionaires at a rate three times that of its nearest competitor (the United States)~\cite{king_2022}, the supply of Chinese-speaking wealth managers has not caught up to this new demand; therefore, each wealth manager serving Chinese high-net-worth individuals carries a disproportionately large number of clients. Hong Kong’s low number of clients per intermediary may reflect the high concentration of wealth managers working there, creating intense competition for clients. In any case, these figures show two things: first, beneficiaries greatly outnumber intermediaries in all cases; second, sanctions on intermediaries would affect many oligarchs at once.

\begin{figure}[!hbt]
    \centering
    \includegraphics[width=0.7\linewidth]{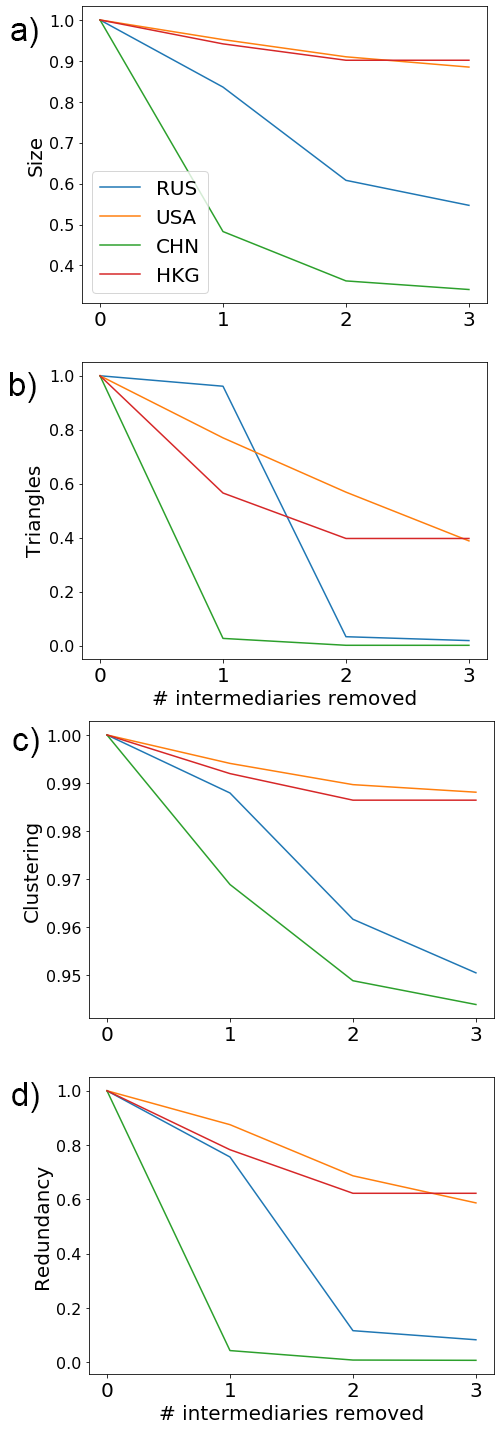}
    \caption{Wealth managers as the points of vulnerability in oligarchs’ offshore financial networks. Noticeably, the offshore networks of high-net-worth individuals from China and Russia are much more vulnerable to such targeted knock-outs of wealth managers than the networks of high-net-worth individuals from the US and Hong Kong.}
    \label{fig:knockout}
\end{figure}

Previous work by Albert et. al.~\cite{albert2000error} shows that “targeted attacks” on the most highly connected nodes can destroy the connectivity of a scale-free network. This property obtains for each of the four nation-level  oligarch networks we construct. However, we also find that Russian and Chinese oligarch networks are characterized by “super fragility” in the face of such attacks. This means that various connectivity markers of the networks collapse completely when a very small number of highly-connected nodes—ties to wealth managers—are removed. Figure~\ref{fig:knockout} shows the results of our iterative removal of the top three wealth managers (by degree) from the offshore networks of high-net-worth individuals from each of the four countries in our study (for the full list of managers, see S2 in the SI). We then assessed global network metrics, normalized on the values from the original network. Specifically, we measured the impact of node removal on a) network size, b) the number of triangles, c) redundancy, and d) the clustering co-efficient. Size refers to the cardinality of the network, which diminishes when nodes are removed; however, if a beneficiary is connected to multiple intermediaries, then removal of one intermediary will not remove the beneficiary. The term triangles refers to the number of triplets that share clients. This tells us how diversified clients are: for instance, if client $a$ and client $b$ both employ wealth managers $x$ and $y$, then even with the removal of $x$ they will remain part of a triangle ($a$--$y$--$b$).  Redundancy measures the total number of pair-wise paths between beneficiaries. Finally, the clustering coefficient represents the fraction of completed triangles out of all possible “triangular” interactions. This is yet another measure of connectivity (see Methods for detailed descriptions). 

The divergent patterns here are quite striking: while the offshore networks of clients from the US and Hong Kong are relatively unperturbed by the removal of intermediaries, the networks of clients from Russia and mainland China collapse after removal of just one or two intermediaries. All four measures—size, triangles, redundancy, and clustering coefficient—diminish significantly after the top intermediary nodes are removed. 

Our findings are consistent with qualitative research on the central role of expert wealth managers in the functioning of the offshore system. In particular, our analyses indicate that these professional intermediaries are the key nodes in the network of global private wealth: they are the point of contact uniting multiple oligarchs and offshore structures across national boundaries~\cite{harrington2015going,hoang2022spiderweb,winters2011oligarchy}. Together, these results suggest why previous sanctions regimes targeting individual oligarchs or offshore jurisdictions have proved easier to bypass than policy-makers expected: they have mistakenly targeted the spokes of a wheel rather than the hub around which the whole system turns.

Figure~\ref{fig:knockout} further suggests that power is not only a matter of concentration and quantity of clients, but of (dis)trust and quality of the client relationship. This is another point suggested by previous qualitative research that we have been able to test and validate through analysis of big data. This aligns with Figure~\ref{fig:powerlaw}(b), which shows Russian and Chinese elites share the secrets of their wealth with a much smaller set of professional intermediaries than do their peers from relatively more free and democratic societies, such as the US and Hong Kong. 

A natural follow-up question is whether degree centrality is the best criterion for knock-out from a disruption viewpoint. Degree centrality is a simple indication of the number of clients an intermediary has. Betweenness centrality, reflecting the number of shortest paths in which a node lies on, is another natural choice and might also be used to target nodes which exhibit high brokerage leverage (in the network scientific term).\footnote{We thank one of the reviewers for this suggestion.}

\begin{figure}[!hbt]
    \centering
    \includegraphics[width=0.8\linewidth]{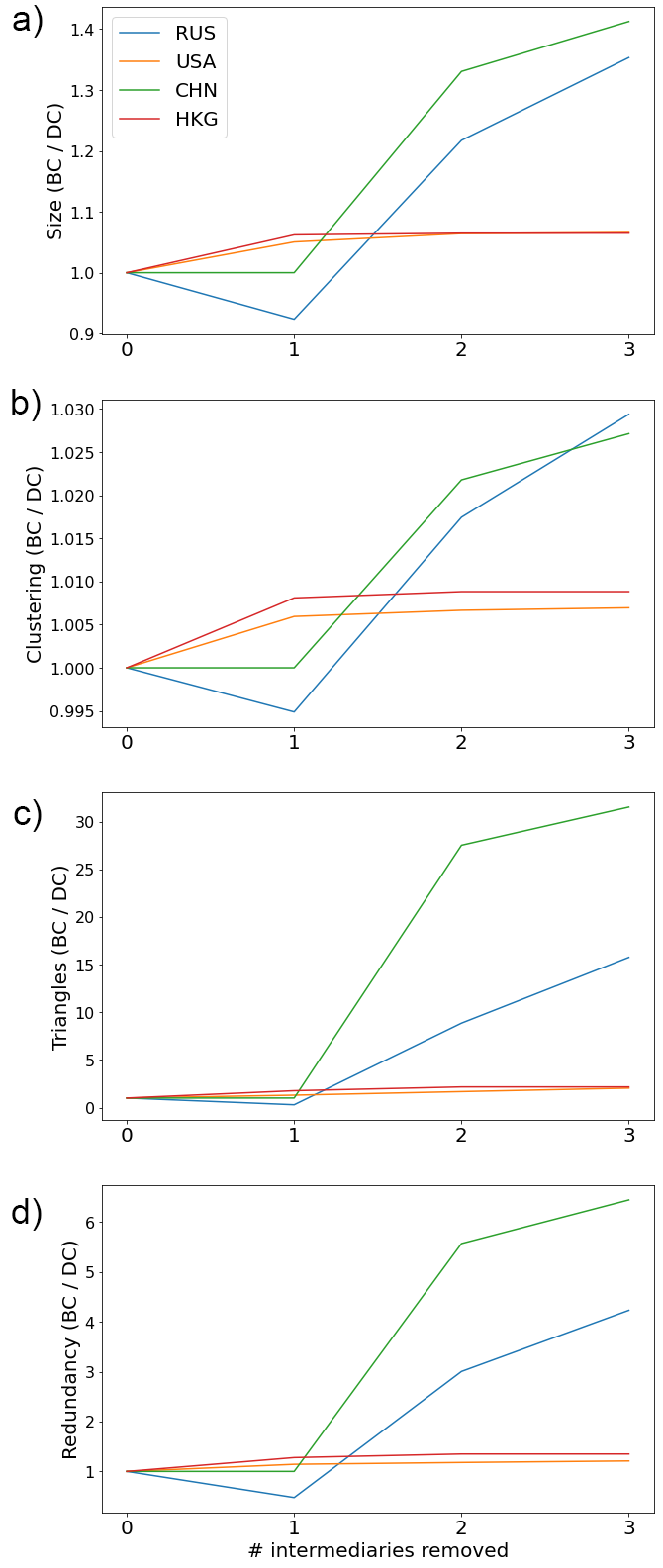}
    \caption{Ratio of knock-out results, based on degree centrality or betweenness centrality. $r>1$ indicates knocknout based on degree centrality is more effective; $r<1$ indicates knockout based on betweenness centrality is more effective. For both USA and HKG, degree-based removal leads to better knockout performance. For CHN and RUS, betweenness-based removal can yield equal or better results, if choosing one.}
    \label{fig:knock-out-bcdc}
\end{figure}

We test this alternative centrality measure by running the same knock-out experiments as before but based on betweenness. For direct comparison purpose, we show the ratio, $r$ of the resulting knock-out impacts by choosing betweenness versus degree centrality, across the four robustness measures. Here, $r > 1$ means greater retention of the disrupted network, hence less impact on the network, and indicates knockout based on betweenness centrality is less effective. On the other hand, $r < 1$ indicates the opposite is true. Figure~\ref{fig:knock-out-bcdc} shows these ratios for the four regions.

For Hong Kong and the US, knockout based on degree centrality performs monotonically better. For China, the first choice in ranking betweenness and degree centrality coincides, which produces a ratio of $r=1$, but the performance of degree centrality-based knockout increases significantly onwards. For Russia however, choosing the intermediary by betweenness centrality to knock out may actually generate a better first response, as the ratio dips below $1$, before increasing above $1$.

This extended result affirms that finding the most crucial nodes in knockout experiments is not a simple matter: the optimal choice is context-dependent. For instance, a common scheme in computer science is modularity maximization~\cite{good2010performance}, which would ensure the maximal knock out of client nodes. However, this does not necessarily correlate with redundancy across clients, which is based on the Cartesian product on other possible paths. 

Thus, in the context of offshore finance, the optimal strategy is likely a weighted combination of two-step betweenness, degree centrality, and mutual nodes across the set of knock-out managers. This is because redundancy is fundamentally defined on both size (how many knocked out) and connectedness (reduction in the product space), and with each country having its own unique configuration. The findings illustrated in Figure~\ref{fig:knockout} and \ref{fig:knock-out-bcdc} represents the ramifications of the scale-free structure of the oligarch networks, however; as indicated by Figure~\ref{fig:powerlaw}(b), there are still important distinctions to be found in the nature of the edge’s spatial relations. 

\begin{figure}[!hbt]
    \centering
    \includegraphics[width=0.7\linewidth]{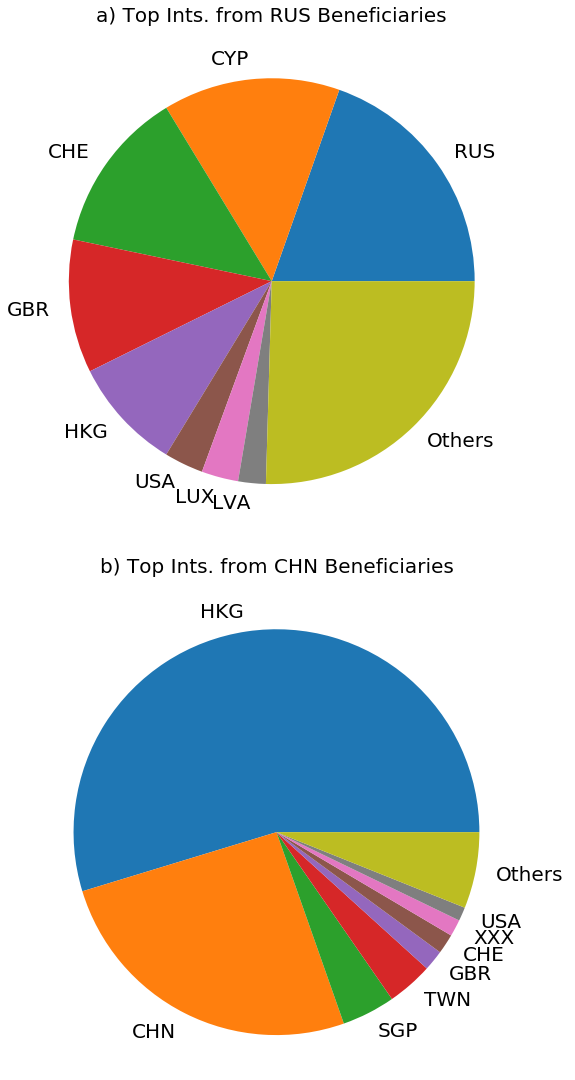}
    \caption{Location of intermediaries for high net-worth clients from (a) Russia versus (b) China. Whereas Russians employ a geographically diverse set of intermediaries, Chinese clients employ wealth managers principally located in Hong Kong. This is equivalent to a diversity index of 0.151 and 0.068 for Russia and China respectively.}
    \label{fig:rus_vs_chn}
\end{figure}

In Figures~\ref{fig:rus_vs_chn}(a) and (b), we drill down into the discrepancy between the power-law and edge quantity findings in the offshore networks of Russian versus Chinese oligarchs. As these Figures show, Russian wealthy elites choose from a far more geographically distributed group of advisors than do the Chinese, who draw their wealth managers primarily from Hong Kong. This is likely due to the aforementioned path dependencies, which—for reasons of British colonial history—put one of the world’s most important offshore financial centers right on China’s doorstep, complete with wealth managers who can speak Chinese and understand the cultural, political and legal concerns of Chinese clients. Russians lack that convenient access portal to the offshore world, and thus by necessity must seek intermediaries elsewhere. This spatial heterogeneity should be acknowledged both by scholars and in future sanctions regimes. Quantitatively, we compute a diversity index (DI) based on the inverse Herfindahl-Hirschman Index (HHI), where higher DI indicates greater diversity (see SI). This yields 0.151 and 0.068 for Russia and China respectively, indicating that Russian oligarchs’ ties to wealth managers are on the order of four times as diverse as those of Chinese oligarchs.

\subsection{Russian Oligarchs Sanctioned Following the 2022 Ukraine Invasion
}

\begin{figure*}[bt]
    \centering
    \includegraphics[width=0.8\linewidth]{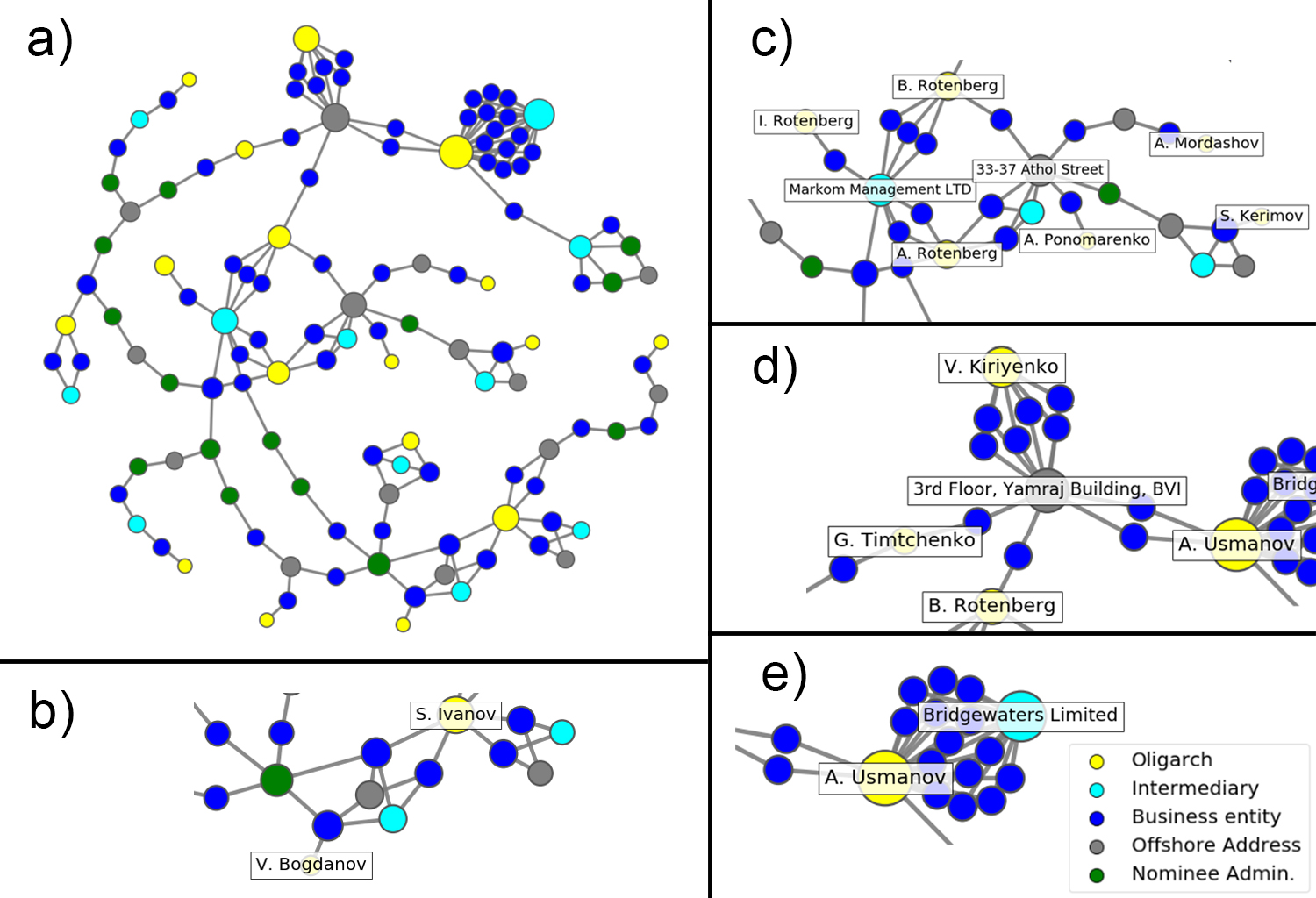}
    \caption{The offshore networks of the 26 sanctioned Russian oligarchs who appear in the ICIJ’s offshore leaks database. These oligarchs are connected through just 124 nodes, forming a rather small social circle of wealth, power and influence. Blue denotes entities, green denotes officers, cyan denotes intermediaries, and yellow the oligarchs.}
    \label{fig:sanctioned}
\end{figure*}

In Figure~\ref{fig:sanctioned}, we isolate the offshore networks of the 26 Russian oligarchs subject to sanctions on their offshore wealth following the February 2022 invasion of Ukraine; these are a subset of Russians sanctioned by the UK, US and EU who also appear in the ICIJ’s offshore leaks database~\cite{icij_2022}. This small group includes some of the wealthiest people in the world, and some of the most powerful in Russia, several of whom have been subject to international sanctions numerous times over the past decade~\cite{tognini_2022}. The relational pattern shown in Figure~\ref{fig:sanctioned}a, which is the induced network on the 26 oligarchs (see Methods), underscores the extreme concentration of these oligarchs’ offshore networks, which can be encapsulated in just 124 nodes.  Note, here we expand our set of nodes to include three additional types. The nodes labeled “Business entity” represent the low-level offshore service providers wealth managers hire to do the paperwork of creating trusts and corporations. The nodes labeled “Nominee Admin” are people who rent their names for use in public documentation, to preserve the privacy of oligarch clients using offshore structures. Finally, nodes labeled “Offshore Address” represent the locations for which trusts and corporations are set up. In these figures, cyan denotes intermediaries and yellow represents the oligarchs; in addition, blue nodes represent entities, green denotes nominees, and grey represents offshore addresses.

Figure~\ref{fig:sanctioned}c shows that intermediary firms such as Markom Management—a boutique wealth management agency based in London—not only link oligarchs to numerous offshore entities, but to other oligarchs. For example, Markom works with three members of the Rotenberg family, who are among the wealthiest people in Russia as well as being close personal friends of President Vladimir Putin. In Figure~\ref{fig:sanctioned}d, we see that Boris Rotenberg is linked via an entity—a company in the British Virgin Islands that does trust services paperwork, presumably hired for Rotenberg by Markom—to Alisher Usmanov, the oligarch “entrusted with servicing financial flows” for Putin’s own offshore wealth. The same pathway links Rotenberg to Vladimir Kiriyenko, son of Putin’s current Deputy Chief of Staff~\cite{partridge_2022}. One might conclude from this that the entities should be sanctioned; but as we know from qualitative research, entities are downstream in the system from wealth managers, and much more numerous~\cite{harrington2016capital}. Entities providing paperwork are easily replaced if sanctioned or eliminated, because the skill set they call upon—compliance with local laws, making sure signatures are in the right places and fees paid to corporate registrars—is readily replicated. Wealth management skills, however, are far more complex, cross-national and not easily substitutable. That’s why oligarchs hire and work with wealth managers, not corporate service providers or nominees. 

Finally, Figure~\ref{fig:sanctioned}e shows the extreme concentration of Alisher Usmanov’s offshore holdings under the auspices of a single Isle of Man-based wealth management firm, Bridgewaters Limited. This illustrates what a “highly-connected node” in a scale-free offshore network looks like in practice. Taken together, the images in Figure~\ref{fig:sanctioned} a--e illustrate that wealth management intermediaries are central not only to Russian oligarchs’ offshore financial networks, but also to their social and professional networks. The intermediaries should therefore be a prime target for sanctions and other efforts aimed at disrupting the activities of those oligarchs. Quantification of these motifs, at a larger scale, is ripe for future research.

\begin{table*}[!htb]
\begin{tabular}{|l|l|l|l|}
\hline
\textbf{Intermediary}               & \textbf{Sanctioned Oligarchs} & \textbf{\# Entities} & \textbf{\# Clients} \\ \hline
Markom Management Ltd.              & 6                             & 140                  & 369                 \\ \hline
American Corporate Services, Inc.   & 4                             & 311                  & 540                 \\ \hline
Appleby Trust (Isle of Man) Limited & 4                             & 458                  & 3,566               \\ \hline
I\&T Consulting Ltd.                & 3                             & 138                  & 483                 \\ \hline
G.S.L. Law \& Consulting            & 3                             & 2,097                & 3,487               \\ \hline
Dr. K. Chrysostomides \& Co.        & 2                             & 69                   & 207                 \\ \hline
Ryon Ltd.                           & 2                             & 305                  & 620                 \\ \hline
Consulco International Ltd.         & 2                             & 3,168                & 5,499               \\ \hline
Appleby Services (Bermuda) Ltd.     & 2                             & 3,645                & 72,316              \\ \hline
Dietrich, Baumgartner \& Partner    & 1                             & 41                   & 162                 \\ \hline
Bridgewater Limited                 & 1                             & 200                  & 666                 \\ \hline
Christodoulos Vassiliades           & 1                             & 287                  & 562                 \\ \hline
Andersen Business Services, Inc.    & 1                             & 521                  & 1,350               \\ \hline
Lotus Holding Company Limited       & 1                             & 1,218                & 2,797               \\ \hline
\end{tabular} \caption{Intermediary firms ordered by the total number of sanctioned Russian oligarchs they serve. From this limited sample, there is a negative correlation between the number of sanctioned oligarchs and each firm’s total number of clients overall, suggesting the special vulnerability of Russian oligarchs’ offshore networks to interventions focused on their wealth managers.} \label{tab:top_intermediaries}
\end{table*}

Table~\ref{tab:top_intermediaries} provides more detail on these concentrated local networks (see Methods). Sorted by the number of sanctioned oligarchs served by each management firm, Table~\ref{tab:top_intermediaries} shows an inverse relationship between the number of sanctioned oligarchs and the firms’ total number of clients. In other words, the more a wealth management firm serves sanctioned Russian oligarchs, the fewer clients that firm has in total. This supports earlier qualitative research which suggested that oligarch-intermediary relationships were characterized by an unusual, almost familial level of intimacy quite unusual for professional work, but common in criminal networks such as those of Mafia families~\cite{harrington2016capital}.

Table~\ref{tab:top_intermediaries} shows that the Bermuda-based multinational offshore law firm Appleby represents the largest number of clients overall in the offshore leaks database; however, they only represent two sanctioned Russian clients, Kerimov and Arkady Rotenberg. This may be because large multinationals serving a wide range of clients—in Appleby’s case including the Queen of England —are careful about protecting their own reputations through exclusion of clients who might be tainted by involvement in corrupt or criminal activities. 

In contrast, smaller boutique firms such as Markom and American Corporate Services Inc--which represent a vanishingly small number of clients compared to Appleby’s--service a disproportionately large segment of the Russian oligarch market while taking on a much smaller client base overall. From the intermediaries’ perspective, this suggests a corporate strategy focused on charging higher fees to legally and reputationally risky clients.  From the perspective of Russian oligarchs, this illustrates an extreme concentration or localization of trust in the hands of a tiny number of professionals who serve clusters of closely tied family and friendship groups. This pattern helps explain why Russian elites’ offshore networks would be so vulnerable to targeted attacks on their intermediaries (Fig 3).

\section{Discussion}
Our findings suggest that the offshore financial system operates as a scale-free network, both globally and on the level of nations, and thus shares the same type of structural vulnerabilities as other scale-free networks, such as the Internet. They are robust in some respects, enduring through random deletion, but some are “super fragile” (in terms of the hallmarks of connectivity) to targeted node removal. Most connectivity is concentrated in a few nodes: in the case of Russian oligarchs’ offshore networks, the high-connectivity nodes consist of a small group of wealth managers. The network structure we have uncovered can explain why earlier rounds of sanctions directed at those oligarchs—the sparsely connected nodes in our analysis—could largely be evaded by some prominent individuals, despite direct seizure of some of their assets.  Our findings suggest that future sanctions should be directed at the professional intermediaries who construct and maintain the offshore system for the oligarchs’ benefit. We identify these expert advisors as overlooked chokepoints in the global financial network. 

To obtain these results, we used data from the offshore leaks databases created by ICIJ from the Panama, Paradise and Pandora Papers leaks to conduct a network attack analysis. We show that for Russian oligarchs—as well as their peers in China—removal of just one or two intermediaries collapsed their offshore networks entirely. The fragility of these networks is not a surprise: offshore finance is by definition an elite phenomenon involving a few hundred thousand ultra-high-net-worth individuals---including just under 2,700 billionaires~\cite{hernandez_2022}---served by an even smaller group of intermediaries~\cite{chavagneux2013tax,harrington2016capital,piketty2021income}. But the distinctive “super fragility” of some offshore networks is a novel finding. Our analysis pinpoints the greatest vulnerabilities in those networks, and identifies important national variations in their structures, which we expect may be influenced by geography, history, language, and governance. 
		
Like any study of elite phenomena, our research is limited by data that can only provide fragmentary snapshots of a system that is observably far larger than anyone has measured. This is a perennial problem with research on elites, and particularly on elite wealth: information is extremely scarce because it is intentionally shrouded in multiple levels of secrecy~\cite{harrington2016capital,hoang2022spiderweb}. Thus, data points are few and representativeness difficult to assess. Our analysis drew from what is now the best data available: the 6.94 terabytes of data and 46.8 million records comprising the Panama, Paradise and Pandora Papers leaks. We were fortunate to have such information, since so much previous research on offshore wealth has had to rely on indirect or imputed measures~\cite{farrell2022weak}. Still, there is much more to learn and our findings should be put to the test against the offshore data likely to emerge in future. A potential limitation of our study is that some in network science have questioned the generalizability of scale-free networks~\cite{holme2019rare}. Fortunately, our results are robust regardless of whether the offshore network is scale-free; as long as the network structure is characterized by extreme heterogeneity in the density of intermediary node connectivity, our findings hold. Our goodness-of-fit appears sufficiently robust, but the broader take-away is that two characteristics uncovered by our analysis—heavy-tail behavior and preferential attachment—align closely with the findings of previous qualitative research, forming a coherent, consistent picture of oligarchs’ offshore networks. 

For scholars of inequality, globalization and finance, our work offers several new insights that may be fruitful for future research. First, our findings substantiate the implications of previous qualitative research~\cite{grigoropoulou2022data} suggesting that professional intermediaries are pivotal in linking and maintaining the offshore financial system, and thus in producing some of its most harmful outcomes, such as exploding wealth inequality and elite corruption~\cite{harrington2016capital,hoang2022spiderweb,winters2011oligarchy}. There should therefore be more social scientific attention to such actors. 

Second, our study could be particularly useful in building out an emerging theory of “complex secrets,” which posits that secretive phenomena like financial corruption have proved robust and resistant to change due to certain patterns in evolved structures that enable the distribution and flow of information~\cite{rilinger2019corporate}. Our analysis of ties between oligarchs and their wealth managers not only represents an empirical case of such a  complex system  of secrecy, reified in a network structure, but suggests that such systems may be characterized by the specific network topology we identify. Our results diverge from prior findings on criminal networks~\cite{morselli2010assessing} showing that the most sensitive nodes are characterized by high betweenness and low degree. Specifically, we find that if restricted to just one intermediary, choosing by betweenness centrality may yield greater disruption to the underlying financial networks of oligarchs from autocratic countries. However, in general, knock-out based on intermediaries' degree centrality is more effective. 

Both sets of results suggest that the information structures underlying complex systems of secrecy are diverse and context-dependent in patterned ways; future research should explore these patterns in greater depth. Even in the settings in which scale-free structure is observed  our experiments show that their points of fragility may require new measures for identification. As we find the set of most sensitive nodes depends on a country's unique network configuration, this suggests exciting new directions of inquiry focused on pinpointing optimal knock-out targets, contingent on characteristics of the clients' country of origin. We look forward to extending these analyses, adding texture to our understanding of the adaptations and variations from which secrecy networks emerge.

As a practical matter, the results of our  attack analysis indicate that future sanctions strategies against rogue states should consider targeting the intermediaries serving the country’s elite, instead of or in addition to sanctions on the individual beneficiaries themselves. We may soon have data to test this claim. In early June, the US banned the provision of some offshore expertise to Russian oligarchs—including accountancy and corporate formation, which are crucial for creating and maintaining the structures that shroud oligarchs' assets in secrecy~\cite{amato_2022}. Shortly thereafter, the EU and the UK implemented similar measures, banning international law and tax experts from those jurisdictions from serving sanctioned Russian oligarchs~\cite{EU_2022,foregin_office_2022,thomas_histed_costelloe_kang_2022}. These bans are backed by penalties including imprisonment and huge fines.\footnote{In the US, current law imposes civil penalties on individual intermediaries up to \$250,000; for criminal convictions, the penalty is up to \$1 million in fines and 20 years in prison (see 50 USC 1705 at https://uscode.house.gov/). 
In the UK and EU, such bans on the provision of expert intermediary services are characterized by the same basic purpose and mechanisms, as well as similar penalties.} 

Given our analysis showing Russian elites' reliance on intermediaries based in the UK (and its dependent territories) and in Cyprus (an EU country), these bans could represent a meaningful blow to the oligarchs’ ability to access their troves of offshore wealth or move them beyond the reach of sanctioning countries. Crucially, the ban on the provision of intermediary services is separate from and more extensive than freezing or seizing any particular asset; rather, the ban means cutting off the expertise pipeline linking sanctioned individuals to their offshore wealth. It’s a more encompassing punishment than losing access to a specific bank account or yacht or private jet.

This approach to sanctions via bans on the provision of expert professional services has been around for decades, and derives its legal basis from defense ministry regulations. Such bans were typically imposed on the transfer of data deemed important to national security and technology related to nuclear, chemical and biological weaponry~\cite{mcgowan2007between}. That such rules are now being applied to financial, legal, and accounting expertise indicates growing recognition by policy practitioners that offshore wealth management can threaten international security and stability. This is consistent with our analysis illustrating the significance of these professionals as chokepoints in global financial networks. Our work thus also motivates a closer look at country-specific sanctions for more policy-oriented outcomes, and inspires subsequent work to take up more nation-level policy comparisons~\cite{siegel2022oligarchs,gould2018economic}.

Most importantly, sanctioning professional intermediaries has a solid track record of effectiveness in achieving policy goals. For example, the current US policy vis-a-vis intermediaries and Russian oligarchs was preceded by a similar ban on intermediaries working with government officials in Iran–which the US sanctions as “State Sponsor of Terrorism.” This approach has been credited with “contributing to Iran’s decision to enter into a 2015 agreement that put limits on its nuclear program,” in part because the country’s leadership “could not access its foreign exchange assets held abroad”\cite{Katzman2022CRS}.

Finally, the highly tailored impact of this strategy makes it far less morally fraught than broad-based sanctions which can deprive whole nations of resources like grain, medical supplies, and fuel. While there is no known data on the frequency with which bans on elite intermediary services are used as part of sanctions packages, this strategy is certainly less well-known publicly than resource sanctions (despite the occasional breathless media coverage of luxury asset seizures). We hope that our study will contribute to more widespread knowledge, discussion, and application of this more precisely targeted tool of international policy.

\section{Methods}
\subsection*{Dataset}
In its natural form, each entry in the IJIP database can be one of five node classes:
\begin{itemize}
    \item \textbf{Officers --} Individual people who are related to entities. We further split this into three classes.\textbf{Beneficiaries:} The direct recipients and benefits of offshore accounts; \textbf{Nominees}: Individuals who are instated to manage these accounts; \textbf{(Officer) Intermediaries:}  Those employed to set-up these accounts.
    \item \textbf{Entities --} The businesses set-up for beneficiaries. These can be corporations, foundations, or trusts.
    \item \textbf{Intermediaries --} Companies that help set-up these offshore entities.
    \item \textbf{Addresses -- }Addresses registered to the three other types of nodes named above.
\end{itemize}
	
These four node classes are then structured through an edge list. Relational in nature, the general structure is as followed: directed edges from officers to entities, intermediaries to entities, and from officers, entities, and intermediaries to addresses. In this study, we focus especially on the tripartite structure of officers, entities, and intermediaries, and use this tripartite network to induce a bipartite network between intermediaries and officers (for a visualization of the tripartite structure, see Figure S1 in the SI). We further filter on the officer class, classifying manually the top fifty most common classes (i.e. shareholder, ultimate beneficiary, nominee executive). The top 50 classes account for 99.2\% of all entries in the relational edge list (see Table S1 in the SI).

\subsection*{Country-level comparisons}
To assess the country-level behavior of beneficiaries, we first filter on beneficiary officer nodes, then extract their connected entities. From the edge list, we then extract the intermediaries connected to these entities.

\begin{equation}
\begin{aligned}
    B_c = \{b \in B \quad | \quad CNTRY(b) = c \} \\\
    E_c = \{e \in E \quad | \quad e \in nei(b) \forall b \in B_c \} \\\
    I_c  = \{i \in I \quad | \quad I \in nei(e) \forall e \in E_c\}
\end{aligned}
\end{equation}

Note, it is possible to subset $B_c$, $E_c$, and $I_c$ directly from the log-level data. Upon acquiring the set of intermediaries tied to beneficiaries from country $c$, we first consider the ratio of these two sets: $\frac{B_c}{I_c}$. We then consider the degree distribution of $I_c$ for each country, which is then plotted as a power law.

\subsection*{Node-removal Experiment}
The purpose of this analysis is to assess the robustness of financial networks in general. We choose the United States, China, and Russia for comparison, adding in Hong Kong due to its role in facilitating Chinese investments and known role as a hub for intermediaries. For each country $c$, we identify the top 3 frequently occurring intermediaries in $I_c$. We then iteratively remove $i_1, i_2,$, and $i_3$.

At each iteration, we assess the size, clustering coefficient, number of triangles, and redundancy. Size denotes the cardinality of the set of nodes, which serves as one view of redundancy in these networks, since if a client only invests through one intermediary, they will be removed. The clustering coefficient is given as:

\begin{equation}
    C = \frac{\text{Num. of closed triangles}}{\text{Num. of possible triangles}}
\end{equation}

As such, the number of triangles provides a more raw and less scaled measure of such redundancy. Intuitively, if clients tend to diversify their intermediaries, more triangles will be present. 
Lastly, our own measure of redundancy counts the possible paths between any two beneficiaries for a given country. We approximate this by taking the square sum of the cardinality of all connected components. Formally, let CC(k) denote the set of connected components when $k$ intermediaries are removed. Then the normalized redundancy metric $R$ is given by:

\begin{equation}
    R = \frac{\Sigma_{a \in CC(i)}|a|(|a|-1) }{\Sigma_{b \in CC(0)}|b|(|b|-1)}
\end{equation}

For size, clustering coefficient, and triangles we use the built-in NetworkX package for direct computation.

\subsection{Sanctioned Oligarch-level Network}
Using fuzzy-word matching, we extracted 26 of 41 sanctioned oligarchs. We then constructed an undirected subgraph between all the 26 oligarchs, restricted to one degree away (up to one direct intermediary). This yielded connected components, for which we computed the shortest paths between these connected components via oligarchs.

Figure~\ref{fig:sanctioned} shows how the sanctioned oligarchs are directly connected and allow us to observe the minimal paths among these clusters of oligarchs. Given the set of intermediaries linked to the Russian Oligarchs (which we denote as $I_{RO}$), we further expanded the network by one degree to tabulate the number of entities and clients each of these intermediaries serves, which gives the power of intermediaries for their sanctioned clients as shown in Table~\ref{tab:top_intermediaries}.

Although our analysis is based on the first wave of sanctions—as of March 1st—we do not expect the network structures and patterns we discovered to change with subsequent waves of sanctions~\cite{butler_neate_boffey_rushe_2022}. We leave this an empirical question for future research. 

\bibliography{refs}

\end{document}